\documentclass[final,a4paper]{aipproc}

\layoutstyle{8x11double}

\usepackage{amsmath,amsfonts}
\usepackage{epsfig,multicol,bbm,array}

\newcommand{\ud}{\mathrm{d}}
\newcommand{\ve}{\varepsilon}
\newcommand{\be}{\begin{equation}}
\newcommand{\ee}{\end{equation}}

\begin{document}

\title{Dynamics of the Bianchi I model with non-minimally coupled scalar field
near the singularity}

\classification{04.50.Kd, 95.36.+x}
\keywords{non-minimal coupling, dynamical dark energy, anisotropy}

\author{Orest Hrycyna}
{ 	
	address={Theoretical Physics Division, National Centre for Nuclear
	Research, Ho{\.z}a 69, 00-681 Warszawa, Poland}
}

\author{Marek Szyd{\l}owski}
{
	address={Astronomical Observatory, Jagiellonian University, Orla 171,
	30-244 Krak{\'o}w, Poland}, 
	altaddress={Mark Kac Complex Systems Research Centre, Jagiellonian
	University, Reymonta 4, 30-059 Krak{\'o}w, Poland}
}

\begin{abstract}
Dynamical systems methods are used to study the evolution of the Bianchi I model with a
scalar field. We show that inclusion of the non-minimal coupling term between the
scalar field and the curvature changes evolution of the model compared with the
minimally coupled case. In the model with the non-minimally coupled scalar field
there is a new type of singularity dominated by the non-minimal coupling term.
We examine the impact of the non-minimal coupling on the anisotropy evolution and
demonstrate the existence of its minimal value in the generic case.
\end{abstract}

\maketitle

\section{Introduction}

In the simplest cosmological model with a scalar field one can introduce a term
modelling its coupling with the curvature. Such a term appears naturally if we
understand the General Relativity as an effective field theory of gravity
\cite{Donoghue:1994dn}. To answer the question how stable effects of the non-minimal
coupling are, we consider the simplest anisotropic space of Bianchi I type which
represents the simplest flat model with anisotropy.

In the model under consideration we assume a universe filled with a non-minimally
coupled scalar field with the curvature and an unknown coupling constant $\xi$.
The action assumes the following form
\begin{equation}
\begin{split}
        S= &\frac{1}{2\kappa^{2}}\int\ud^{4}x\sqrt{-g}R \\ &
        -\frac{1}{2}\int\ud^{4}x\sqrt{-g}
        \left\{\ve\nabla^{\alpha}\phi\,\nabla_{\alpha}\phi
	+ \ve\xi R\phi^{2}+
        2U(\phi)\right\}
\end{split}
\label{eq:action}
\end{equation}
where $\kappa^{2}=8\pi G$, $\ve=\pm1$ corresponds to a canonical and phantom
scalar field, respectively, the metric signature is $(-,+,+,+)$ and $U(\phi)$
is the scalar field potential function assumed in a polynomial form.

\section{Bianchi I metric}
We assume the following form of the metric
\begin{equation}
	\ud s^{2} = -\ud t^{2} + a^{2}(t)\Big(g_{1}^{2}(t)\ud x^{2} +
	g_{2}^{2}(t)\ud	y^{2} + g_{3}^{2}(t)\ud z^{2}\Big)\,,
\label{eq:bmetric}
\end{equation}
together with the condition
\begin{equation}
g_{1}g_{2}g_{3}=1\,.
\label{eq:metcon}
\end{equation}

The energy conservation condition we obtain from variation of the action
(\ref{eq:action}) with respect to the metric components
\begin{equation}
\begin{split}
	3H^{2}-\Sigma = \kappa^{2}\Big( &\ve \frac{1}{2}\dot{\phi}^{2} +
			      U(\phi) \\ &
+\ve\xi(3H^{2}-\Sigma)\phi^{2}+\ve3\xi H (\phi^{2})\dot{}\Big)
\end{split}
\label{eq:energy}
\end{equation}
where the anisotropy is measured by
\begin{equation}
\begin{split}
	\Sigma & =\frac{1}{2}\left(\left(\frac{\dot{g_{1}}}{g_{1}}\right)^{2} +
	\left(\frac{\dot{g_{2}}}{g_{2}}\right)^{2} +
	\left(\frac{\dot{g_{3}}}{g_{3}}\right)^{2}\right) \\ 
	& =\frac{1}{2}\left(q_{1}^{2} + q_{2}^{2} + q_{3}^{2}\right)\,,
\end{split}
\label{eq:anib}
\end{equation}
the Hubble function is given by $H=\frac{\dot{a}}{a}$ and the space volume is
$V=a^{3}$.

The acceleration equation is
\begin{equation}
\begin{split}
	\dot{H}=&-2 H^{2} -\frac{1}{3}\Sigma + \\ & +
	\frac{\kappa^{2}}{6}\frac{-\ve(1-6\xi)\dot{\phi}^{2}+4U(\phi)-6\xi\phi
U'(\phi)}{1-\ve\xi(1-6\xi)\kappa^{2}\phi^{2}}\,.
\end{split}
\end{equation}
From space-space Einstein's equations we have additional equations of motion for
anisotropy
\begin{equation}
(1-\ve\xi\kappa^{2}\phi)\dot{q_{i}} = \ve\xi\kappa^{2} q_{i}(\phi^{2})\dot{}-3H
q_{i}(1-\ve\xi\kappa^{2}\phi^{2})\,.
\label{eq:animot}
\end{equation}

When the non-minimal coupling is present one can write the right hand sides of Einstein field equations
in several possible inequivalent ways \cite{Faraoni:book}. In the case adopted here the energy
momentum tensor of the scalar field is covariantly conserved, which may not be
true for different approaches. For example, one can redefine the gravitational
constant $\kappa_{\text{eff}}^{-2} = \kappa^{-2}-\ve\xi\phi^{2}$ making it time
dependent, then the effective gravitational constant can diverge for a critical
value of the scalar field $\phi_{c}=\pm1/\sqrt{\ve\kappa^{2}\xi}$ and the model
is unstable there with respect to arbitrary small anisotropy perturbations which
become infinite there \cite{Starobinsky:1981}.

In what follows we introduce the energetic (expansion normalised) variables
$$
x\equiv\frac{\kappa}{\sqrt{6}}\frac{\dot{\phi}}{H}\,,\quad
y\equiv\frac{\kappa}{\sqrt{3}}\frac{\sqrt{U_{0}}}{H}\,,\quad
z\equiv\frac{\kappa}{\sqrt{6}}\phi\,,\quad \Omega_{\Sigma} =
\frac{\Sigma}{3H^{2}}\,,
$$
and $U(\phi) \to f(z)$, then the energy conservation condition can be expressed as
\begin{equation}
\begin{split}
	1 = & \hspace{1mm} y^{2}f(z) + \ve(1-6\xi)x^{2} + \ve6\xi(x+z)^{2} +\\ & + 
	(1-\ve6\xi z^{2})\Omega_{\Sigma}\,,
\end{split}
\label{eq:consb}
\end{equation}
and the acceleration equation 
\begin{equation}
	\begin{split}
		\frac{\dot{H}}{H^{2}} = & -2 -\Omega_{\Sigma} - \\
		& -\frac{\ve(1-6\xi)x^{2}-y^{2}\big(2f(z)-3\xi z
	f'(z)\big)}{1-\ve6\xi(1-6\xi)z^{2}}\,.
	\end{split}
\label{eq:accelb}
\end{equation}
The equations of motion for the anisotropy are
\begin{equation}
	(1-\ve6\xi z^{2})\frac{1}{H}\frac{\dot{q_{i}}}{q_{i}} =\ve12\xi x z
	-3(1-\ve6\xi z^{2})
\label{eq:anisomotion}
\end{equation}
Note that for the minimally coupled scalar field $\xi=0$ equations
(\ref{eq:animot}) and (\ref{eq:anisomotion}) can be directly integrated and the
evolution of the anisotropy does not depend whether the universe is filled with
a canonical or phantom scalar field. In general case of the non-minimal coupling
this is not possible because these equations depend on the phase space variables.

The dynamical system is in the following form
\begin{equation}
\begin{split}
	\frac{\ud x}{\ud \ln{a}} & = -x(1-\Omega_{\Sigma})
    -\ve\frac{1}{2}y^{2}f'(z)-\\ & \quad-(x+6\xi
	z)\Big(\frac{\dot{H}}{H^{2}}+2+\Omega_{\Sigma}\Big)\,,\\
	\frac{\ud y}{\ud \ln{a}} & = -y\frac{\dot{H}}{H^{2}}\,,\\
	\frac{\ud z}{\ud \ln{a}} & = x\,.
\end{split}
\label{eq:dynsysb}
\end{equation}
where using equations (\ref{eq:consb}) and (\ref{eq:accelb}) we eliminate
$\Omega_{\Sigma}$ and $\dot{H}/H^{2}$. One can see that right hand sides of the
dynamical system (\ref{eq:dynsysb}) are rational functions of their arguments.
In order to use standard dynamical systems methods we need to make the following
time transformation 
\begin{equation}
	(1-\ve6\xi
	z^{2})(1-\ve6\xi(1-6\xi)z^{2})\frac{\ud}{\ud\ln{a}}=\frac{\ud}{\ud\tau}\,
\label{eq:timerep}
\end{equation}
which removes singularities of the system at $z^{2}=\frac{1}{\ve6\xi}$ and
$z^{2}=\frac{1}{\ve6\xi(1-6\xi)}$.

In this work we concentrate our attention at the following critical point $$(x^{*}=-6\xi z^{*},\quad y^{*}=0,\quad
(z^{*})^{2}=\frac{1}{\ve6\xi(1-6\xi)})$$ which for the isotropic case corresponds to
the finite scale factor singularity \cite{Hrycyna:2010yv}. The existence of this
critical point depends on the value of the coupling constant $\xi$ only. For the
canonical scalar field $\ve=+1$ it exists for $0<\xi<\frac{1}{6}$ while for the
phantom scalar field $\ve=-1$ for $\xi<0$ or $\xi<\frac{1}{6}$.

Eigenvalues of the linearization matrix at this critical point are
$$
\lambda_{1,2}=-\frac{36 \xi^{2}}{1-6\xi}\,,\quad \lambda_{3}=
-\frac{72\xi^{2}}{1-6\xi}\,.
$$
Stability, in time $\tau$, of this critical point depends on the value of the coupling constant
$\xi$. For $\xi<\frac{1}{6}$ the eigenvalues are negative indicating a stable critical
point, while for $\xi>\frac{1}{6}$ the eigenvalues are positive which indicates an
unstable critical point. 

The solutions of the linearised system in the vicinity of the critical point
under investigation are
\begin{equation}
	\begin{split}
		x(\tau) &= x^{*} + \big(\Delta x+2(1-3\xi)\Delta
		z\big)\exp{(\lambda_{1}\tau)} - \\ & \quad- 2(1-3\xi)\Delta z
		\exp{(\lambda_{3}\tau)}\,\\
		y(\tau) & = y^{*} + \Delta y \exp{(\lambda_{2}\tau)}\,,\\
		z(\tau) & = z^{*} + \Delta z \exp{(\lambda_{3}\tau)}\,.
	\end{split}
\end{equation}
where $\Delta x = x^{(i)} - x^{*}$, $\Delta y = y^{(i)}-y^{*}$ and $\Delta z =
z^{(i)}- z^{*}$ are displacements of the initial conditions with respect to
coordinates of the critical point.

From (\ref{eq:consb}) we see that at this critical point
$\Omega_{\Sigma}^{*}=0$, and using the linearised solutions we have
\begin{equation}
	\Omega_{\Sigma}(\tau) = \ve2(1-6\xi)^{2} z^{*} \Delta z
	\exp{(\lambda_{3}\tau)}\,.
\end{equation}
On the other hand the anisotropy parameter $\Sigma=3
H^{2} \Omega_{\Sigma}$ and in the energy phase space variables
$\frac{1}{\kappa^{2}U_{0}}\Sigma=\frac{\Omega_{\Sigma}}{y^{2}}$.
Using linearised solutions in the vicinity of this critical point
we have 
\begin{equation}
\frac{1}{\kappa^{2}U_{0}}\Sigma(\tau)=\frac{\Omega_{\Sigma}(\tau)}{(y(\tau))^{2}}
=\ve2(1-6\xi)^{2} z^{*}\frac{\Delta z}{(\Delta y)^{2}}=\text{const.}
\end{equation}
As long as the linear solutions are considered, one can see, that the anisotropy
function is strictly constant and the value of anisotropy in the vicinity of the
critical point representing the finite scale factor singularity depends on the
value of the non-minimal coupling constant $\xi$ and initial conditions in the linear
approximation. Of course one can feel bad about the square value of $\Delta y$ in
the denominator which should vanish in the linear approximation, but in the
dynamical system (\ref{eq:dynsysb}) the phase space variable $y$ appears only in
a polynomial form of the second degree and one can make the following change of
variables $\widetilde{y}=y^{2}$ and then the same results hold.

\begin{figure}
\begin{tabular}{c}
	{\includegraphics[scale=0.475]{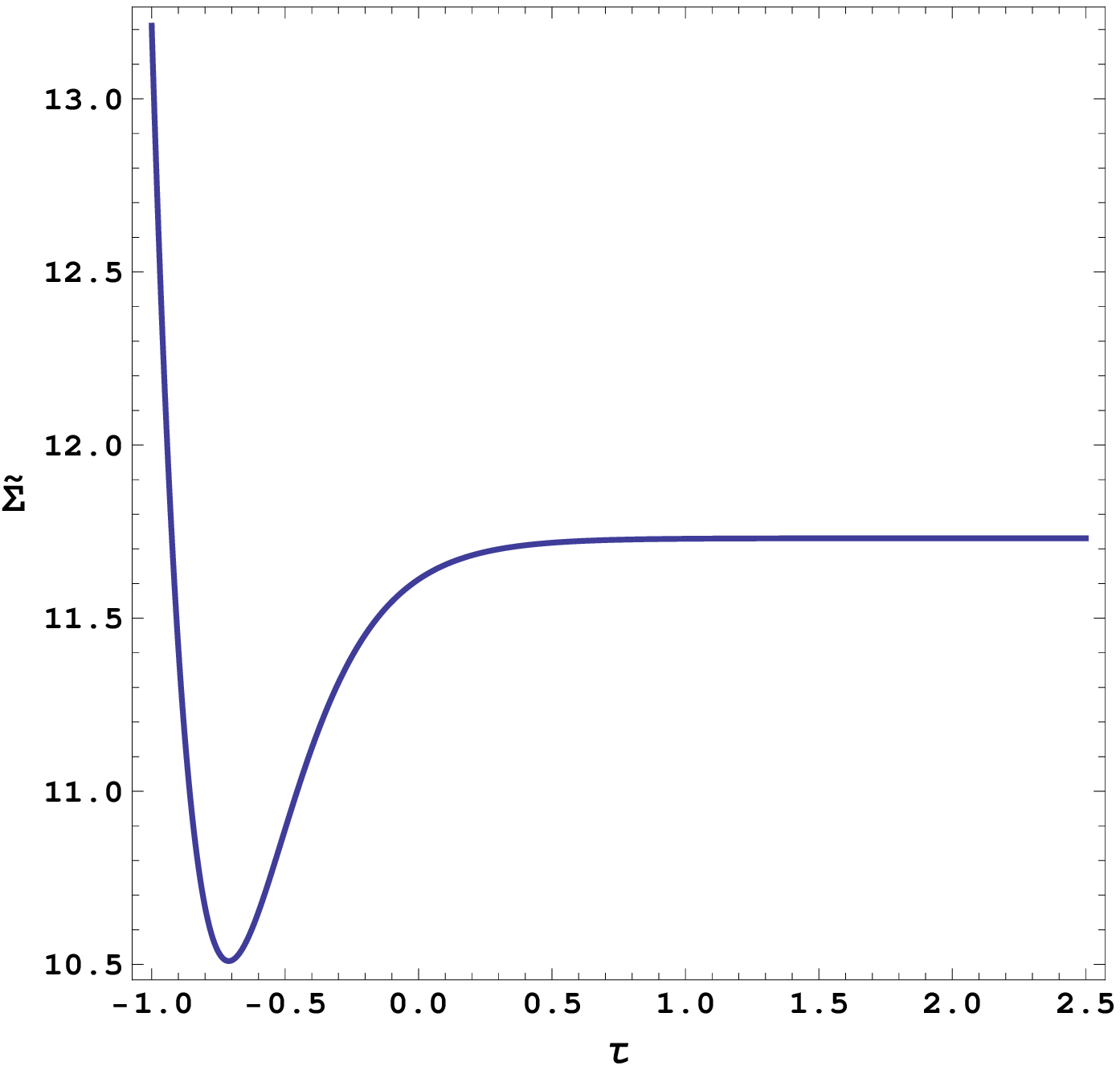}}\\
	{\includegraphics[scale=0.475]{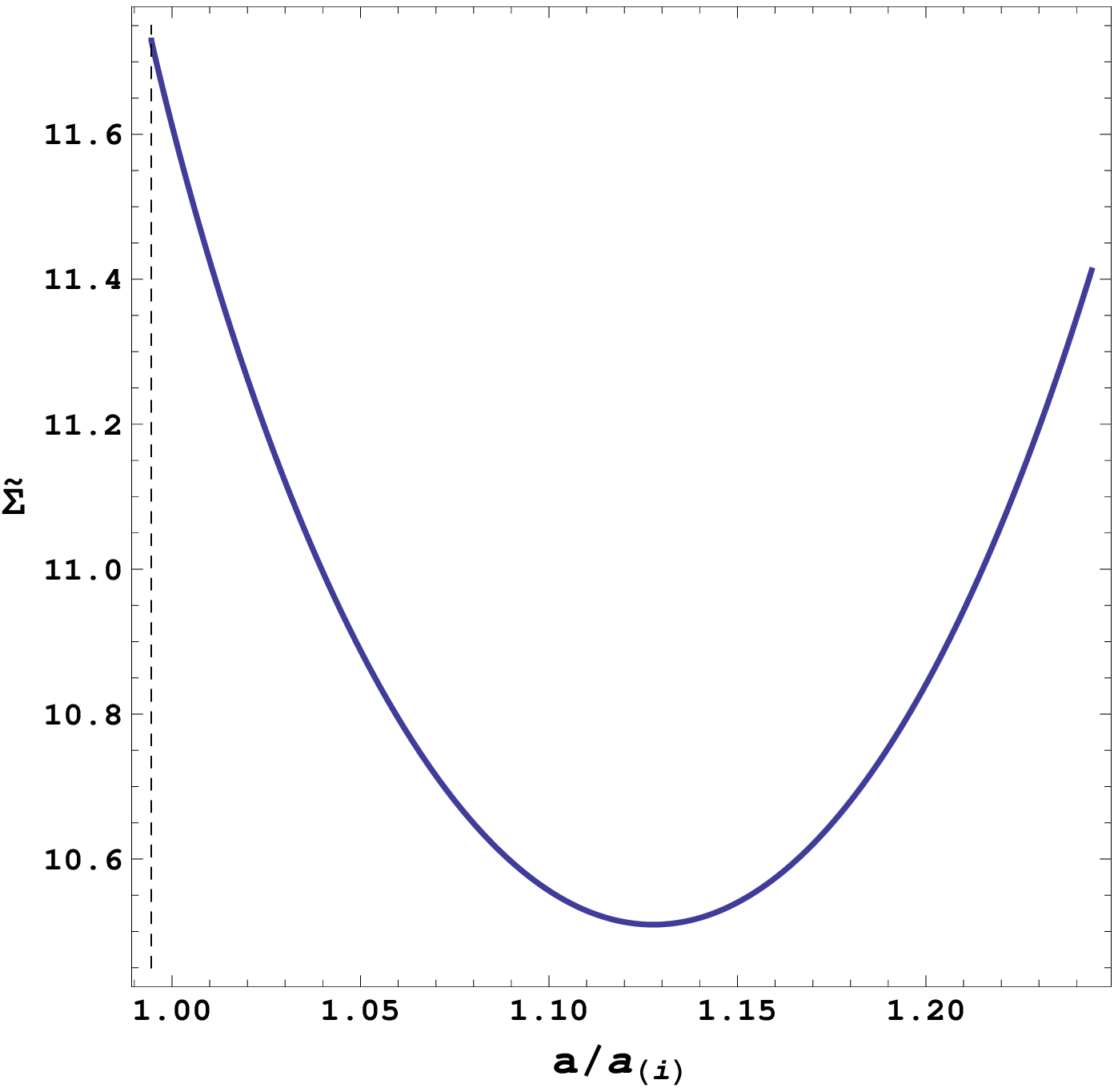}}
\end{tabular}
	\caption{Evolution of the anisotropy function
	$\tilde{\Sigma}=\frac{\Sigma}{\kappa^{2}U_{0}}$ calculated from
(\ref{eq:consb}) and using $3H^{2}=\frac{\kappa^{2}U_{0}}{y^{2}}$. The initial
conditions taken in the vicinity of the critical point under investigations, and
$U(\phi)\propto\phi^{2}$, $\ve=+1$, $\xi=\frac{1}{8}$.
We observe that the anisotropy is constant at the critical point then approaches
a minimal value and next grows as the universe expands. A dashed vertical line
denotes value of the scale factor of the singularity.}
	\label{fig:1}
\end{figure}

In Figure \ref{fig:1} we plotted evolution of the anisotropy. At
the first plot we present evolution of $\tilde{\Sigma}$ with respect to time
$\tau$
and the critical point representing a finite scale factor singularity is
approached at $\tau\to\infty$. At the second plot we present evolution of the
anisotropy with respect to the scale factor $a/a_{(i)}$ rescaled to the value of
the scale factor taken at the initial conditions.

\section{Kasner metric}

In this section in (\ref{eq:bmetric}) we assume
$$
g_{1}=t^{p_{1}}\,, \quad g_{2}=t^{p_{2}}\,,\quad g_{3}=t^{p_{3}}\,,
$$
and the metric is
\begin{equation}
	\ud s^{2} = -\ud t^{2} + a^{2}(t)\Big(t^{2p_{1}}\ud x^{2} +
	t^{2p_{2}}\ud y^{2} + t^{2p_{3}}\ud z^{2}\Big)\,,
\end{equation}
where from (\ref{eq:metcon}) we have constraint equation $$p_{1}+p_{2}+p_{3}=0\,.$$
The assumed form of the metric represents a special case of the general Bianchi I metric which
can be recast into the standard Kasner solution for an empty space.

Now the anisotropy function is
\begin{equation}
\Sigma=\frac{1}{2}(p_{1}^{2}+p_{2}^{2}+p_{3}^{2})\frac{1}{t^{2}}=3\frac{\sigma^{2}}{t^{2}}
\end{equation}
and the energy conservation condition expressed in the energetic variables is
\begin{equation}
\begin{split}
	1= &\hspace{1mm} y^{2}f(z)+\ve(1-6\xi)x^{2}+\ve6\xi(x+z)^{2}+ \\ 
	  &+(1-\ve6\xi
	z^{2})\sigma^{2}h^{2}\,,
\end{split}
\label{eq:consk}
\end{equation}
where we defined
$$
h=\frac{1}{H t}\,.
$$
From (\ref{eq:anisomotion}) we obtain
\begin{equation}
	(1-\ve6\xi z^{2})h=3(1-\ve6\xi z^{2})-\ve12\xi x z\,.
\label{eq:hconstr}
\end{equation}
The dynamical system is in the following form
\begin{equation}
\begin{split}
	\frac{\ud x}{\ud \ln{a}} & = -x(1-\sigma^{2}h^{2})
   -\ve\frac{1}{2}y^{2}f'(z)-\\ & \quad-(x+6\xi
	z)\Big(\frac{\dot{H}}{H^{2}}+2+\sigma^{2}h^{2}\Big)\,,\\
       \frac{\ud y}{\ud \ln{a}} & = -y\frac{\dot{H}}{H^{2}}\,,\\
       \frac{\ud z}{\ud \ln{a}} & = x\,.
\end{split}
\end{equation}
where using equations (\ref{eq:consk}) and (\ref{eq:accelb}) (after substitution
$\Omega_{\Sigma}=\sigma^{2}h^{2}$) we
eliminate terms containing $\sigma^{2}h^{2}$ and $\dot{H}/H^{2}$.
The dynamical system describing evolution in the Kasner metric is exactly in the
same form as in the case of the general Bianchi I model. One exception is the existence
of the additional constraint equation (\ref{eq:hconstr}). Using this equation we
find that at the critical point under investigations $h^{*}=1$ this indicates
that while a sample trajectory approaches a critical point and $H\to\infty$ then
$t\to0$.
\begin{figure}
	\begin{tabular}{c}
		{\includegraphics[scale=0.475]{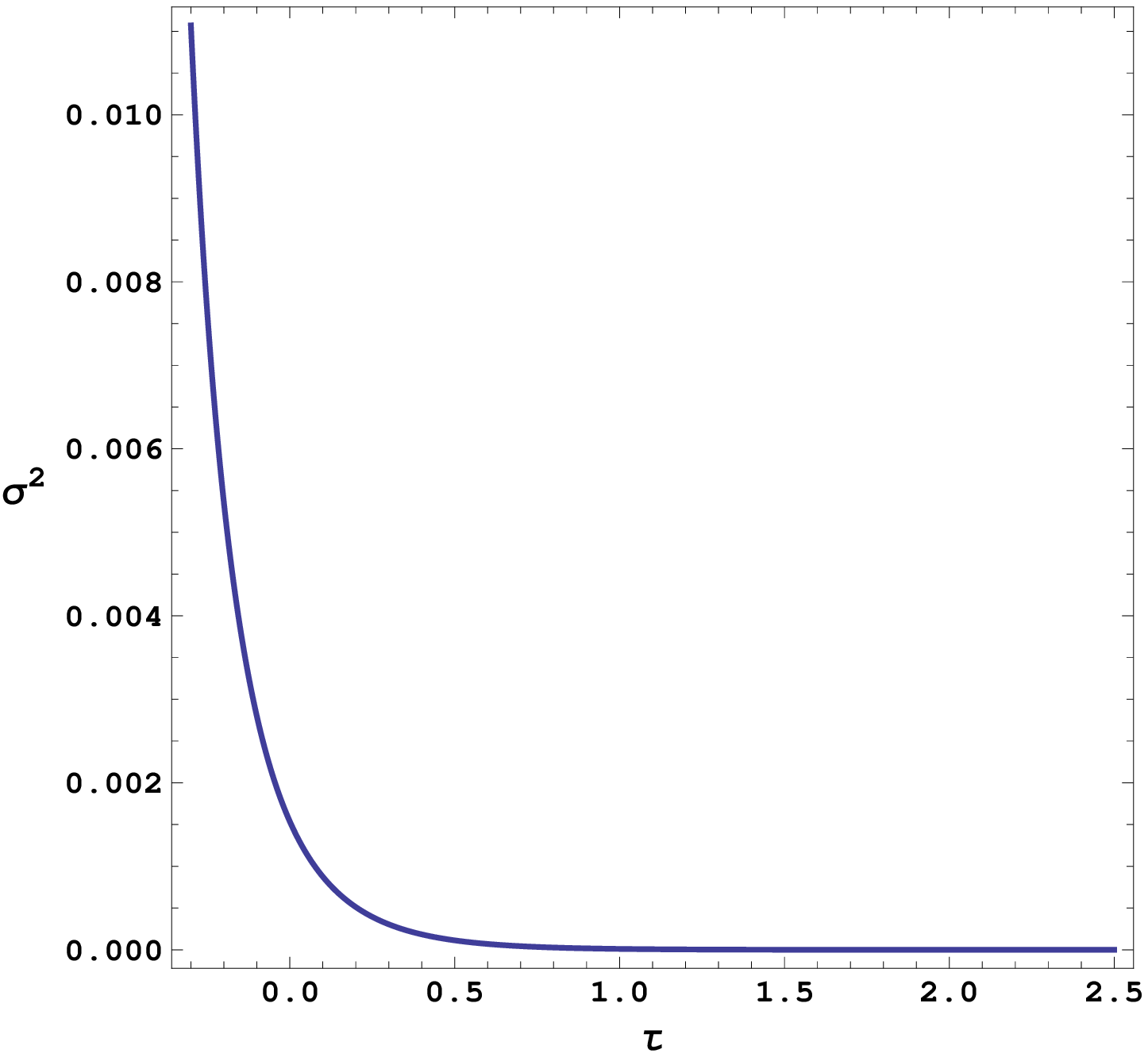}}\\
		{\includegraphics[scale=0.475]{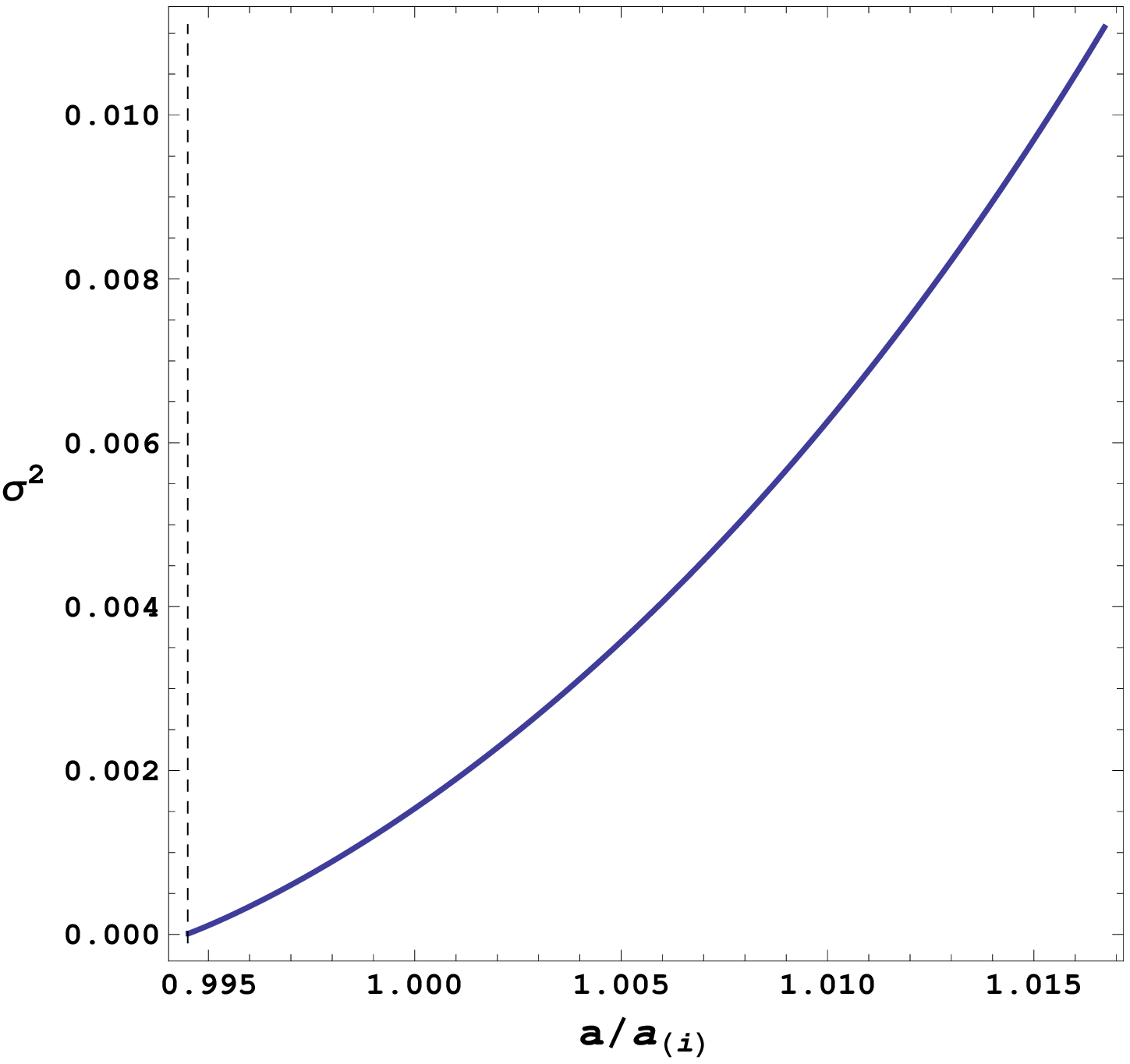}}
	\end{tabular}
\caption{Evolution of the anisotropy parameter $\sigma^{2}$ calculated from
	(\ref{eq:sigma}), and $U(\phi)\propto\phi^{2}$, 
	$\ve=+1$, $\xi=\frac{1}{8}$. In the Kasner metric anisotropy vanishes at the critical point
representing the finite scale factor singularity. As the universe expands anisotropy
grows indicating that this critical point is unstable with respect to anisotropy
perturbations. A dashed vertical line denotes value of the scale factor of the
singularity.}
\label{fig:2}
\end{figure}

The energy conservation condition (\ref{eq:consk}) can be rewritten into
the following form
\begin{equation}
\sigma^{2}=\frac{1-y^{2}f(z)-\ve(1-6\xi)x^{2}-\ve6\xi(x+z)^{2}}{(1-\ve6\xi
z^{2})h^{2}}
\label{eq:sigma}
\end{equation}
where $h$ is given by (\ref{eq:hconstr}) and using the linearised solutions
obtained in the previous section we find
\begin{equation}
	\sigma^{2} = \ve2(1-6\xi)^{2} z^{*} \Delta z \exp{(\lambda_{3}\tau)}\,.
\end{equation}

In Figure \ref{fig:2} we plotted evolution of the anisotropy parameter
$\sigma^{2}$ for the Kasner metric. In evolution in time $\tau$ (first plot)
critical point under investigations is approached as $\tau\to\infty$. At the
second plot we presented evolution of $\sigma^{2}$ with respect to the scale
factor $a/a_{(i)}$. Both plots show that during the contraction of universe the
anisotropy for the Kasner metric decreases and vanishes at the finite scale
factor singularity.

\section{Conclusions}

Dynamics of the simple anisotropic cosmological model with non-minimally
coupled scalar field was investigated near the singularity. The critical point
representing the finite scale factor singularity exists only for non-minimal
$\xi\ne0$ and non-conformal $\xi\ne\frac{1}{6}$ values of the coupling constant.
We have shown that for the most general form of the metric anisotropy of the
universe reaches minimal value and then model starts to anisotropies. Moreover in
the finite scale factor singularity the value of the anisotropy tends to a
constant value which depends on the non-minimal coupling. On the other hand,
if we consider only small anisotropies, the universe isotropises during the
contracting phase and reaches an isotropic flat cosmological model in the finite
scale factor singularity. As the universe expands small anisotropies grow
indicating instability with respect to arbitrary small anisotropy perturbations.

\begin{theacknowledgments}
The research of OH was funded by the National Science Centre
through the post-doctoral internships award (Decision No.
DEC-2012/04/S/ST9/00020).
\end{theacknowledgments}

\bibliographystyle{aipproc}

\bibliography{../darkenergy,../standard}

\begin{thebibliography}{4}
\expandafter\ifx\csname natexlab\endcsname\relax\def\natexlab#1{#1}\fi
\providecommand{\enquote}[1]{``#1''}
\expandafter\ifx\csname url\endcsname\relax
  \def\url#1{\texttt{#1}}\fi
\expandafter\ifx\csname urlprefix\endcsname\relax\def\urlprefix{URL }\fi
\providecommand{\eprint}[2][]{\url{#2}}

\bibitem[Donoghue(1994)]{Donoghue:1994dn}
J.~F. Donoghue, \emph{Phys.Rev.} \textbf{D50}, 3874--3888 (1994),
  \eprint{gr-qc/9405057}.

\bibitem[Faraoni(2004)]{Faraoni:book}
V.~Faraoni, \emph{{Cosmology in Scalar--Tensor Gravity}}, vol. 139 of
  \emph{Fundamental Theories of Phsics}, Kluwer Academic Publishers,
  Dordrecht/Boston/London, 2004.

\bibitem[Starobinsky(1981)]{Starobinsky:1981}
A.~A. Starobinsky, \emph{Sov. Astron. Lett.} \textbf{7}, 36--38 (1981).

\bibitem[Hrycyna and Szydlowski(2010)]{Hrycyna:2010yv}
O.~Hrycyna, and M.~Szydlowski, \emph{JCAP} \textbf{12}, 016 (2010),
  \eprint{1008.1432}.

\end{thebibliography}

\end{document}